\documentclass[manuscript]{aastex}

\usepackage{color}

\shorttitle{Relativistic Electrons in the DG Tau Jet}
\shortauthors{Ainsworth et al.}

\begin{document}

\title{Tentative Evidence for Relativistic Electrons Generated by the \\ Jet of the Young Sun-like Star DG~Tau}

\author{Rachael E. Ainsworth\altaffilmark{1}}
\affil{Dublin Institute for Advanced Studies, 31 Fitzwilliam Place, Dublin 2, Ireland}
\email{rainsworth@cp.dias.ie}

\author{Anna M. M. Scaife}
\affil{School of Physics \& Astronomy, University of Southampton, \\ Highfield, Southampton, SO17 1BJ, UK}

\author{Tom P. Ray\altaffilmark{1}, Andrew M. Taylor} 
\affil{Dublin Institute for Advanced Studies, 31 Fitzwilliam Place, Dublin 2, Ireland}


\author{David A. Green and Jane V. Buckle\altaffilmark{2}}
\affil{Cavendish Laboratory, J J Thomson Avenue, Cambridge, CB3 0HE, UK}

\altaffiltext{1}{School of Physics, Trinity College, Dublin 2, Ireland}
\altaffiltext{2}{Kavli Institute for Cosmology, Madingley Road, Cambridge, CB3 0HA, UK}

\begin{abstract}
Synchrotron emission has recently been detected in the jet of a massive protostar, providing further evidence that certain jet formation characteristics for young stars are similar to those found for highly relativistic jets from AGN. We present data at 325 and 610\,MHz taken with the GMRT of the young, low-mass star DG~Tau, an analog of the Sun soon after its birth. This is the first investigation of a low-mass YSO at such low frequencies. We detect emission with a synchrotron spectral index in the proximity of the DG~Tau jet and interpret this emission as a prominent bow shock associated with this outflow. This result provides tentative evidence for the acceleration of particles to relativistic energies due to the shock impact of this otherwise very low-power jet against the ambient medium. We calculate the equipartition magnetic field strength $B_{\rm min}\approx0.11$\,mG and particle energy $E_{\rm min}\approx4\times10^{40}$\,erg, which are the minimum requirements to account for the synchrotron emission of the DG~Tau bow shock. These results suggest the possibility of low energy cosmic rays being generated by young Sun-like stars.
\end{abstract}


\section{Introduction}
\label{sec:intro}

Jets are powered by astrophysical objects spanning an extreme range of masses, from young brown dwarfs ($<10^{-2}$\,M$_{\odot}$) and young stellar objects (YSOs, $\sim1$\,M$_{\odot}$) to Galactic microquasars and active galactic nuclei (AGN, $\sim10^{8}$\,M$_{\odot}$). In spite of the enormous differences in scale, their similar morphologies and other characteristics suggest that the launching mechanism may be universal \citep{2011AIPC.1358..329L}, although this suggestion is still controversial. However, it is widely accepted that many astrophysical jets, including those from YSOs, are driven by accretion and launched through magnetohydrodynamic processes, in particular the centrifugal acceleration of material along magnetic field lines anchored to a rotating disk \citep{1982MNRAS.199..883B,2006A&A...453..785F}. In the case of YSOs, the rotation of the circumstellar disk, and therefore of the magnetic field lines, creates a helical configuration that is believed to collimate the jet via ``hoop stresses'' occurring tens of au from the protostar. The role of the magnetic field on larger scales, however, is not well understood \citep{2009RMxAC..36..179R} and direct observational support for its existence is still lacking.

Synchrotron radiation has been detected in the jet of a massive YSO using linear polarization measurements from the Very Large Array (VLA), allowing for the direct measurement of the strength and direction of the magnetic fields within the young system \citep[HH~80-81,][]{2010Sci...330.1209C}. This is a remarkable result as the presence of relativistic particles was previously thought to be unlikely in jets from young stars where the velocities are only of the order of hundreds of km\,s$^{-1}$, in comparison with jets from AGN where the jet velocities are close to the speed of light. Many features of the observed polarization in HH~80-81 are similar to those seen in more highly energetic AGN jets, strengthening the argument that certain jet formation characteristics span huge mass and energy scales. 

The radio emission at gigahertz frequencies from \textit{low}-mass YSOs is typically detected as thermal bremsstrahlung radiation \citep[e.g.][]{1996ASPC...93....3A} emitted by their partially ionized, collimated outflows \citep{1986ApJ...304..713R}. The free--free spectrum from such emission is characterized by a power-law spectral index $\alpha$, where the flux density $S_{\nu}\propto\nu^{\alpha}$ at frequency $\nu$ and varies from $-0.1\lesssim\alpha\lesssim2$. Non-thermal emission is known to arise from the more evolved and jet-less Class~III YSOs \citep{1992A&A...259..149W}, however radio observations for a handful of low-mass Class~0-II protostars also imply non-thermal processes may be at work, such as gyrosynchrotron emission \citep{1997Natur.385..415R}, suggesting the presence of at least mildly relativistic electrons spiraling along magnetic field lines. In contrast to that of free--free emission, the synchrotron spectrum has $\alpha<-0.5$ which, together with the brightness, relates to the energy spectrum of the electrons. 

DG~Tau is one of the most enigmatic young stars studied today. It is a highly active T~Tauri star located at a distance of 140\,pc in the Taurus Molecular Cloud (TMC), and has a bipolar outflow \citep[e.g.][]{1997A&A...327..671L} although the blueshifted side is better studied. It was one of the first T~Tauri stars to be associated with an optical jet and exhibits several shocks and knots in its outflow, the most prominent of which was first observed $8''$ away from the star \citep[HH~158,][]{1983ApJ...274L..83M}. The optical blueshifted jet has a position angle (P.A.) of $223^{\circ}$ close to the source \citep{1997A&A...327..671L}, however large-scale studies trace the jet out to a total projected distance of 0.5\,pc and show that the P.A. of the outflow changes to $218^{\circ}$ \citep[HH~702,][]{2004A&A...420..975M}. Additionally, knots and clumps in the outflow have proper motions that deviate from the bulk jet motion \citep{1998AJ....115.1554E}. In the radio, DG~Tau has been shown to have a compact and elongated morphology in the known outflow direction and to possess a spectral index typical of free--free emission at frequencies $>5$\,GHz \citep{1982ApJ...253..707C,1986AJ.....92.1396C,2012A&A...537A.123R,2012MNRAS.420.3334S,2013ApJ...766...53L,2013MNRAS.436L..64A}. 

In this letter we present deep radio continuum observations at 325 and 610\,MHz of DG~Tau using the Giant Metrewave Radio Telescope \citep[GMRT,][]{2005ICRC...10..125A} located near Pune, India. \textbf{This is the first investigation of a low-mass YSO at sub-gigahertz frequencies}. In Section~2 we list the details of the observations, present our results in Section~3, and make our concluding remarks in Section~4.

\section{Observations and data reduction}
\label{sec:obs}

Observations of DG~Tau $(\alpha, \delta)_{\rm J2000.0}=(04^{\rm h}27^{\rm m}04.\!^{\rm s}7, +26^{\circ}06'15''\!.74)$ \citep{2013AJ....145...44Z} were made using the GMRT at 325 and 610\,MHz between 6--13 December 2012 (average epoch 2012.95). Observations at 325\,MHz were taken for two hours per night over the course of three nights for a total of 6 hours; observations at 610\,MHz were taken for two hours in a single run. A total bandwidth of 32\,MHz was observed, which was split into 256 spectral channels. The sample integration time was 16.9\,s. The primary beam of the GMRT has a width at half power of approximately $81'$ at 325\,MHz and $43'$ at 610\,MHz. 

The flux density scale was set through observations of 3C48 at the beginning and end of each observing run. For the cases where 3C48 set before the end of the run, 3C147 or 3C286 were observed. The flux densities were calculated using the Astronomical Image Processing Software (\textsc{AIPS}) task \textsc{setjy} \citep{2013ApJS..204...19P} and were found to be 45.6\,Jy at 325\,MHz and 29.4\,Jy at 610\,MHz for 3C48, 55.1\,Jy at 325\,MHz for 3C147, and 20.8\,Jy at 610\,MHz for 3C286. The phase was calibrated using interleaved 3\,min observations of J0431+206 every 20\,min. The \textsc{AIPS} task \textsc{getjy} retrieved flux densities of 2.78$\pm$0.02\,Jy at 325\,MHz and 3.05$\pm$0.05\,Jy at 610\,MHz for J0431+206.
 
Flagging of baselines, antennas, channels, and scans that suffered heavily from interference was performed for each night's observation using standard \textsc{AIPS} tasks. Bandpass calibrations for each antenna were obtained from the observations of the flux calibrator source and applied to the data. Five central frequency channels were then combined together with the task \textsc{splat} and antenna-based phase and amplitude calibration was performed with \textsc{calib}. This calibration was then applied to the original data and averaged into 24 separate spectral channels. Target source data was then extracted and concatenated with the data from the additional nights for imaging. 

Due to the large primary beam of the GMRT, we followed \citet{2007MNRAS.376.1251G} and conducted wide-field imaging using facets. Images of each field were produced with \textsc{robust} set to 0 within \textsc{imagr} to optimize the trade-off between angular resolution and sensitivity. However due to only a marginal source detection, we re-imaged the 610\,MHz data convolved with the same \textsc{CLEAN} beam as the 325\,MHz data to better detect any extended emission and directly compare the two datasets. The synthesized beam has dimensions $\Theta_{\rm FWHM} = 11''\!.6\times9''\!.2$, P.A.~$79^{\circ}\!.6$. We recover an rms noise of $\sigma_{\rm rms}=146\,\mu$Jy at 325\,MHz and $\sigma_{\rm rms}=93\,\mu$Jy at 610\,MHz. 

There was a systematic displacement of $2''$ in declination between the $325$ and $610$\,MHz images due to the difference of ionosphere with declination between DG~Tau and the phase calibrator ($\approx6^\circ$ apart). We align the images using the positions of several compact (extragalactic) sources in the field. We find the positions at $610$\,MHz in agreement with the NRAO VLA Sky Survey \citep{1998AJ....115.1693C}, and therefore shift the 325\,MHz image $2''$ to the south. An overlay of the resulting maps is shown in Fig.~\ref{fig1}.

\section{The DG~Tau bow shock}
\label{sec:res1}

\begin{figure*}
\centerline{\includegraphics[width=0.75\textwidth]{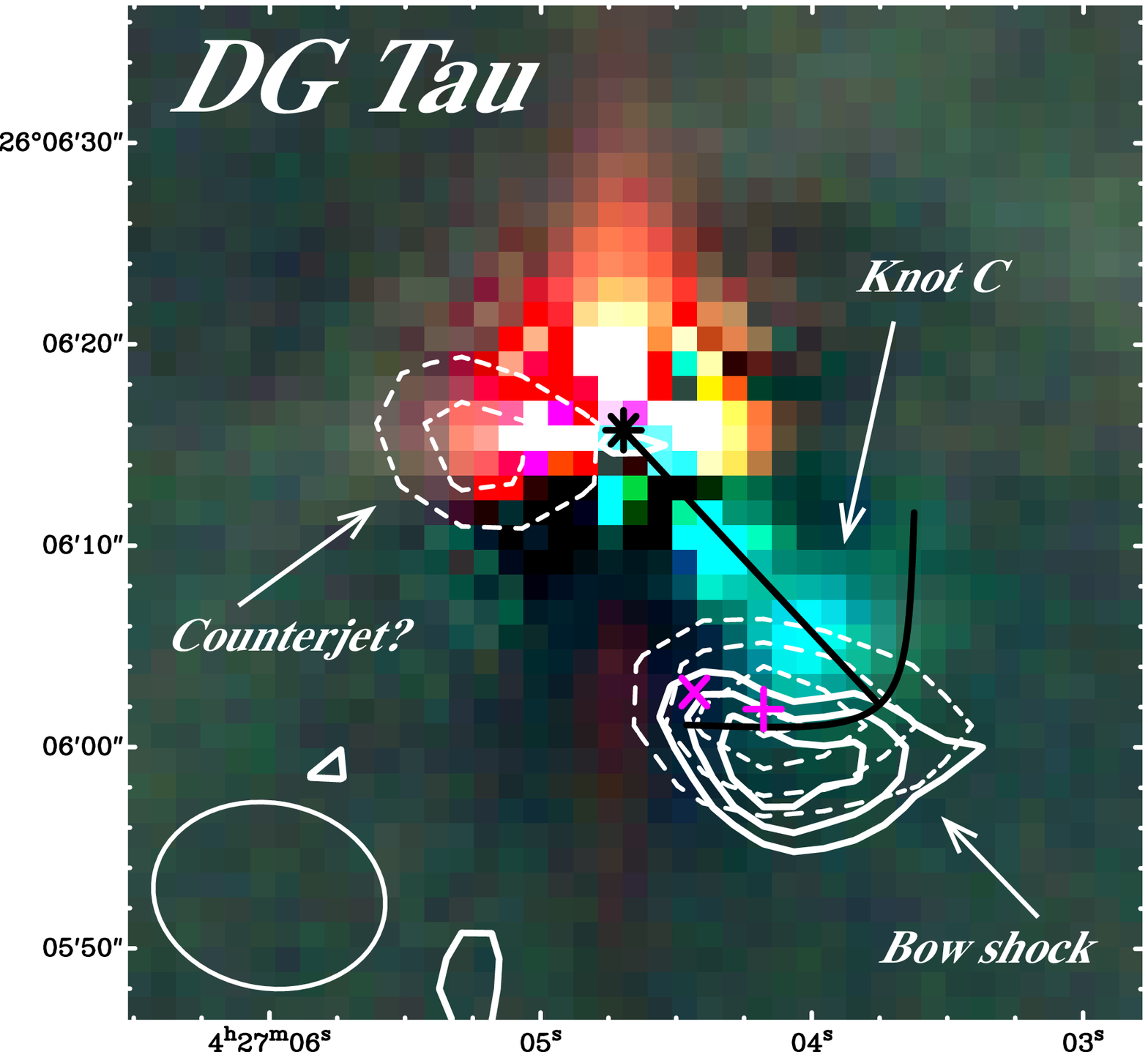}}
\caption{Our GMRT observations at 325\,MHz (dashed contours) and 610\,MHz (solid contours) overlaid on a composite RGB image built from I, H$\alpha$ and [SII] bands from the TLS Schmidt telescope at a similar epoch (2012.92; B. Stecklum, priv. comm.) to illustrate detection of the bow shock driven by Knot~C from \citet{1998AJ....115.1554E}.  
GMRT contours are $-3, 3, 4, 5, 6\times\sigma_{\rm rms}$, where $\sigma_{\rm rms}=146\,\mu$Jy\,beam$^{-1}$ at 325\,MHz and $93\,\mu$Jy\,beam$^{-1}$ at 610\,MHz, although we note there are no negative contours within the section of the field shown. 
The synthesized beam is shown as an ellipse in the bottom left corner. 
All coordinates are J2000.0. 
The EVLA positions of the bow shock at 5.4 and 8.5\,GHz are shown as a plus (+) and a cross ($\times$), respectively \citep[see Table~\ref{tbl1};][]{2013ApJ...766...53L}.
The optical stellar position corrected for proper motion \citep{2013AJ....145...44Z} is shown as an asterisk ($\ast$) and the optical jet axis and bow shock are shown as solid black lines. 
We note there is a $3\sigma$ contour at 610\,MHz at the optical stellar position tracing the base of the jet that may be difficult to see.
 \label{fig1}}
\end{figure*}

Although we marginally detect DG~Tau at $4\,\sigma_{\rm rms}$ at 325\,MHz and $3\,\sigma_{\rm rms}$ at 610\,MHz with the GMRT, the focus of this letter is the detection of the significant extended emission $\approx18''$ (2,500\,au) to the southwest (see Fig.~\ref{fig1}). Discussion on the emission associated with the stellar position, including the offset of the emission between the two frequencies, will be presented in a forthcoming paper. We tentatively suggest that the emission at 325\,MHz near the star is associated with the counterjet, although further investigation is needed.

Based on its curved morphology at 610\,MHz, we identify this radio emission to be associated with the prominent bow shock reported by \citet[][driven by their Knot~C]{1998AJ....115.1554E} when due allowance is made for its tangential motion. These authors detect the associated bright knot $8''\!.7$ away from the star (epoch 1986.99), traveling across the sky at $(0''\!.168\pm0''\!.007)$\,yr$^{-1}$ and P.A.$=216^{\circ}\!.6\pm2^{\circ}\!.7$. Based on a simple extrapolation in time, the position of the knot is expected to be $\approx13''$ away from the star at the epoch of our GMRT observations, and this is confirmed by nearly coeval optical observations taken with the Alfred Jensch Schmidt Telescope (epoch 2012.92, B. Stecklum, private communication). We detect emission at the edge of the optical knot (see Fig.~\ref{fig1}), indicating the emission arises close to, if not at, the shock front itself. 

There is a slight offset of the peak emission from the optical jet axis (and thus apex of the bow shock). Radio emission at this location is also apparent in nearly coeval (epoch 2012.22) observations from the expanded-VLA at 5.4 and (marginally) 8.5\,GHz \citep[][see also \citealt{2012A&A...537A.123R}]{2013ApJ...766...53L}. The peak of the significant signal detected with the EVLA at 5.4\,GHz is in almost perfect agreement with our emission at 325\,MHz and within $\sim3''$ of the peak of the 610\,MHz emission. The positional uncertainty of the bow shock in the GMRT data is $\sim1''$ at both 325 and 610\,MHz based on the spatial resolution and signal-to-noise ratios \citep{1998AJ....115.1693C}, and we find the peak locations of our GMRT data consistent within these uncertainties. 

We extrapolate the results of \citet[][e.g., following similar procedures as in \citealt{1998AJ....116.2953A}]{1984ApJ...287..461C}, and calculate the likelihood of this emission arising from a background extragalactic object based on the separation from DG~Tau and the optical jet axis. The probability of finding a background object within $\pm20^\circ$ in P.A. from the jet axis and within a distance of $18''$ from the star (an area of about $110$\,arcsec$^{2}$), is 0.0004 (or $\sim10^{-3}$ if considering a symmetric bipolar jet). It is therefore reasonable to propose that this emission arises from the DG~Tau bow shock. A plausible explanation for the offset from the optical jet axis could be due to the jet slamming into an inhomogenous medium.

\subsection{A Synchrotron Jet}
\label{sec:res2}

Due to the asymmetric (and therefore non-Gaussian) nature of the emission at 325 and 610\,MHz, we extract flux densities for the bow shock following the technique described in \citet{2007BASI...35...77G} and the results are listed in Table~\ref{tbl1}. The errors are calculated as $\sigma = \sqrt{(0.05S_{\nu})^2+\sigma_{\rm{rms}}^2+\sigma_{\rm{fit}}^2}$, where $0.05S_{\nu}$ is an assumed 5\% absolute calibration error on the flux density $S_{\nu}$, $\sigma_{\rm rms}$ is the rms noise and $\sigma_{\rm fit}$ is the standard deviation of the fluxes measured in the five polygonal apertures. The errors are in practice dominated by $\sigma_{\rm rms}$. 

\begin{figure}
\centerline{\includegraphics[width=\textwidth]{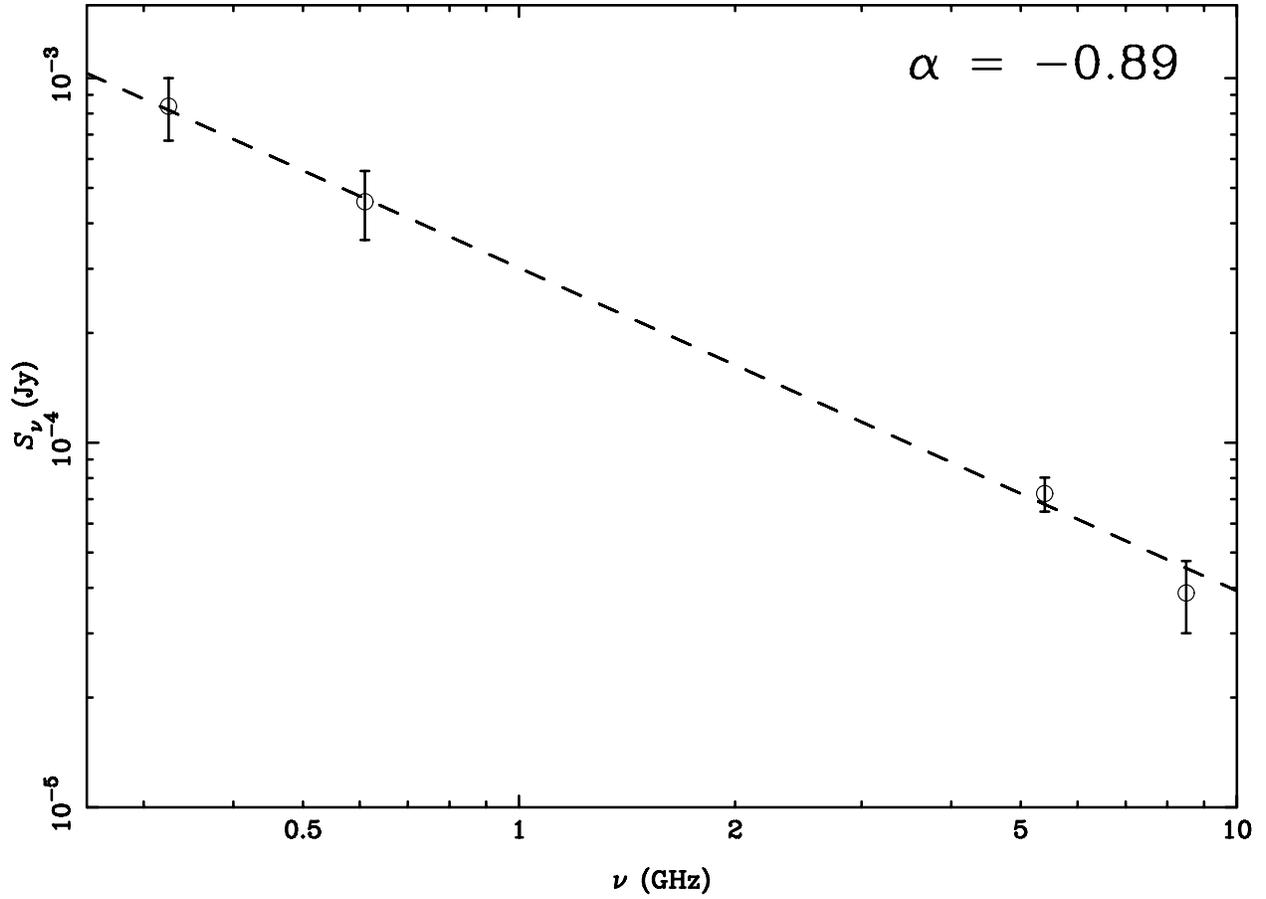}}
\caption{Spectrum of the DG~Tau bow shock. The GMRT data at 325 and 610\,MHz and EVLA data at 5.4 and 8.5\,GHz used are listed in Table~\ref{tbl1}. Dashed line denotes a spectral index $\alpha=-0.89$ (see Section~\ref{sec:res2}). \label{fig2}}
\end{figure}

We extract flux densities from EVLA data at 5.4 and 8.5\,GHz \citep{2013ApJ...766...53L} following the same technique as described above (image files obtained from C. Lynch, private communication) and combine these data with our GMRT measurements to construct the spectrum for the DG~Tau bow shock (see Fig.~\ref{fig2}). Flux densities used in the fitting and positions of the peak emission of the bow shock with frequency are listed in Table~\ref{tbl1}.

Using the Markov Chain Monte Carlo based Maximum Likelihood algorithm {\sc METRO} \citep{Hobson2004} we fit a power-law spectral index $\alpha$ to the spectrum (see application in \citealt{2012MNRAS.423.1089A}) and find $\alpha=-0.89\pm0.07$, distinctly discrepant with free--free emission and indicative of synchrotron radiation. We dismiss gyrosynchrotron radiation as the emission mechanism at this location in the jet as kilogauss magnetic field strengths would be required which are expected to be present only close to the stellar surfaces \citep{1997Natur.385..415R}. 

The combination of the synchrotron spectrum with the detection of the emission at the edge of the optical knot implies active particle acceleration at the shock. The acceleration of relativistic particles may not be altogether unlikely in shocked regions where the fast thermal jet impacts on the ambient medium \citep{2010Sci...330.1209C}. Electrons and protons can gain energy via the first-order Fermi mechanism, by diffusing upstream across a shock front and gaining additional energy after recrossing downstream relative to the shock \citep{1991MNRAS.251..340D}.  

Although the radio spectra of YSOs are typically dominated by thermal radiation emitted by jet plasma, we are beginning to see the dominance of an additional emission mechanism at these low frequencies. This investigation of a young star at very low frequencies exposes a new regime for studying the non-thermal processes, and hence the magnetic fields, associated with YSOs. Although the function of the magnetic field in the role of jet launching has been hotly debated, its relevance on large-scales is less studied and the magnetic field strength remains the last piece of the physical puzzle regarding protostellar jets \citep{2009RMxAC..36..179R}. 

We therefore use our result for the spectral index along with the standard equations given by \citet{2011hea..book.....L} to calculate the minimum requirements to generate the synchrotron emission: the equipartition magnetic field strength,  
\begin{equation}
B_{\rm min} = \left[ \frac{3\mu_{0}}{2} \frac{G(\alpha) (1+k) L_{\nu}}{Vf} \right]^{2/7},
\end{equation}
and energy of the relativistic particles,
\begin{equation}
E_{\rm min} = \frac{7}{6\mu_{0}} (Vf)^{3/7} \left[ \frac{3\mu_{0}}{2} G(\alpha) (1+k) L_{\nu} \right]^{4/7}.
\end{equation}
In these equations, $L_{\nu}$ is the luminosity of the source at frequency $\nu$, $V$ and $f$ are the volume and volume filling factor of the emitting region respectively, $\mu_{0}$ is the vacuum permeability, $k$ is the ratio of the energy of the heavy particle population to that of the electron population, and $G(\alpha)$ is a function of the spectral index and of the minimum and maximum frequencies of the observed radio spectrum.  

For our measured flux density at 610\,MHz, $\alpha=-0.89$, $V\simeq6\times10^{43}$\,m$^{3}$ with $f=0.5$ (see Fig.~\ref{fig1}), and $k=40$ for electrons undergoing Fermi shock acceleration in a non-relativistic jet \citep{2005AN....326..414B}, we find $B_{\rm min}\simeq0.11$\,mG and $E_{\rm min}\simeq4\times10^{40}$\,erg for the DG~Tau bow shock. Such a field strength is similar to values obtained from Zeeman observations in star-forming cores \citep{1999ApJ...520..706C}.

\subsection{Cosmic rays from a low-mass YSO jet?}
\label{sec:res3}

Diffusive shock acceleration is thought to be the mechanism that gives rise to cosmic rays (CRs), and therefore jets from YSOs could potentially provide a new laboratory to test the theory of Galactic CR production \citep{2010Sci...330.1184R}. Shocks with large Mach numbers ($M_1=v_1/c_1\gg1$) may give rise to efficient particle acceleration, where $v_1$ is the shock velocity and $c_1$ is the sound speed. For our $v_1\approx110$\,km\,s$^{-1}$ \citep{1998AJ....115.1554E} and a temperature of $T\sim10^4$\,K, the sound speed at the bow shock is $c_1\approx10$\,km\,s$^{-1}$ and therefore $M_1\approx11$, which satisfies the limit of very strong shocks \citep{2011hea..book.....L}. 

CR electrons accelerated in association with star formation have already been shown to arise from the Galactic Center \citep{2013Natur.493...66C} and parallel studies of gamma rays and radio emission have revealed an excess at GeV energies within the Milky Way \citep{2013NuPhS.239...64O}. Assuming $E_{\rm min}$ is released over the dynamical timescale of the bow shock \citep[$t_{\rm bow}\sim100$ years,][]{1998AJ....115.1554E}, the luminosity released in the form of low energy CR electrons would be $L_{e}\approx E_{\rm min}/(k\,t_{\rm bow}) \sim 3\times10^{29}$\,erg\,s$^{-1}$ over this period. Furthermore, if the average outflow phase of a T~Tauri star lasts $\tau\sim1$\,Myr and the star formation rate in the Milky Way is $\sim1$ star per year, the number of stars in the Galaxy in this phase at any given time is thus $\sim10^{6}$. Therefore, the total Galactic luminosity of CR electrons generated in YSO jets is $L_{e}^{\rm tot.}\sim3\times10^{35}$\,erg\,s$^{-1}$.

Due to their low energy and the strength of the cloud magnetic field, we calculate the built-up energy density in the cloud to compare with that in the ISM, as it is not clear whether these CR electrons can escape the molecular cloud region over its lifetime. The TMC has an approximate volume $V_{\rm c}\sim3\times10^{53}$\,m$^{3}$ \citep{2010ApJS..186..259R}, therefore the output from a single YSO gives rise to an energy density of $U_{e} = W_{e}/V_{\rm c} = L_{e}\tau/V_{\rm c} \approx 2\times10^{-5}$\,eV\,cm$^{-3}$. Assuming an average cloud lifetime of 10\,Myr \citep[e.g.][]{2007ARA&A..45..339B} and a star formation rate of $\sim8\times10^{-5}$\,M$_{\odot}$\,yr$^{-1}$ \citep[forming YSOs of mass $\sim1$\,M$_{\odot}$;][]{2010ApJ...724..687L}, an energy density of $U_{e}\approx2\times10^{-2}$\,eV\,cm$^{-3}$ can be supported. We note that the Galactic CR electron energy density is $\sim10^{-2}$\,eV\,cm$^{-3}$, hence the pressure from CR electrons accelerated by these YSO systems within the cloud is similar to that from the ISM. This calculation has been done for a region with a known low star formation efficiency and the corresponding energy density for regions of higher efficiency, for example in regions such as the core of the Orion Nebula \citep[$\sim7\times10^{-4}$\,M$_{\odot}$\,yr$^{-1}$,][]{2010ApJ...724..687L}, could be considerably higher. It is therefore plausible that CR electrons could escape from molecular cloud regions over their lifetimes.

From our result for $B_{\rm min}$ and the photon energy $E_{\gamma}=2.5\times10^{-6}$\,eV for $\nu=610$\,MHz, we calculate a Lorentz factor of $\Gamma\approx1400$, which yields an electron energy of 700\,MeV. For this $\Gamma$ we find a Larmor radius of $r_{\rm L}\approx2\times10^{-3}$\,au and therefore an \textit{e}-folding time of approximately 18 days. Such a small \textit{e}-folding time implies ongoing acceleration at the shock, although factors such as turbulence can considerably increase this timescale.

\section{Conclusions}

We have presented the first investigation of a low-mass YSO at sub-gigahertz frequencies and our results suggest that non-thermal processes may begin to dominate the radio spectra of protostellar jets in this regime. We detect emission with a synchrotron spectral index in the proximity of the DG~Tau jet and interpret this emission as a prominent bow shock associated with this outflow. This result provides tentative evidence for the acceleration of particles to relativistic energies due to the shock impact of this otherwise very low-power jet against the ambient medium. We calculate the equipartition magnetic field strength ($B_{\rm min}\approx0.11$\,mG) and particle energy ($E_{\rm min}\approx4\times10^{40}$\,erg) necessary to account for the synchrotron emission from the DG~Tau bow shock. Although follow-up observations are required to obtain the proper motion of the radio emission to verify its origin in the jet and detect linearly polarized emission to measure the direction of the magnetic field, these results suggest the possibility of low energy cosmic rays being generated by young Sun-like stars.

\acknowledgments

We thank the anonymous referee for their careful reading of the manuscript and constructive feedback that helped to clarify these results. We thank the staff of the GMRT that made these observations possible. GMRT is run by the National Centre for Radio Astrophysics of the Tata Institute of Fundamental Research. We thank Christene Lynch of the University of Iowa for providing the EVLA FITS files for position and flux measurements, and Bringfried Stecklum of Th\"{u}ringer Landessternwarte Tautenburg for providing the optical FITS files for verification of the knot location. REA and TPR acknowledge support from Science Foundation Ireland under grant 11/RFP/AST3331. AMS gratefully acknowledges support from the European Research Council under grant  ERC-2012-StG-307215 LODESTONE.

{\it Facilities:} \facility{GMRT}.

\clearpage

\clearpage

\clearpage

\begin{table}
\begin{center}
\caption{DG~Tau bow shock results.\label{tbl1}}
\begin{tabular}{ccccc}
\tableline\tableline
Telescope	&	$\nu$				& $S_{\nu}$ 		& 	$\alpha_{J2000.0}$ & $\delta_{J2000.0}$ 	  \\
		&	(MHz)  				& ($\mu$Jy)		& 	($^{\rm h}~~^{\rm m}~~^{\rm s}$) & ($^{\circ}~~{\arcmin}$~~\arcsec) \\
\tableline
GMRT	&	325					&	$837\pm163$	&	04 27 04.15		&	+26 06 01.3 \\
GMRT	&	610					&	$458\pm98$	&	04 27 04.18		&	+26 05 59.3 \\
EVLA	&	5400\tablenotemark{a}	&	$73\pm8$		&	04 27 04.18		&	+26 06 01.9 \\
EVLA	&	8500\tablenotemark{a}	&	$39\pm9$		&	04 27 04.44		&	+26 06 02.7 \\
\tableline
\end{tabular}
\tablecomments{Column [1] lists the telescope used, [2] the observing frequency, [3] the integrated flux density, and columns [4] and [5] list the Right Ascension and Declination of the peak emission measured with the AIPS task \textsc{maxfit}.}
\tablenotetext{a}{The EVLA observations at 5.4 and 8.5\,GHz are from \citet{2013ApJ...766...53L}, see Section~\ref{sec:res2} for details.}
\end{center}
\end{table}

\end{document}